\documentclass[twocolumn,showpacs,amsmath,amssymb]{revtex4}
\usepackage{graphicx}
\usepackage{bm}
\begin{document}
\title{Gapless Fermi Surfaces in anisotropic multiband superconductors in magnetic field.}
\author{Victor Barzykin} 
\affiliation{Department of Physics and Astronomy, University of Tennessee,
Knoxville, TN  37996-1200}
\author{L. P. Gor'kov}
\altaffiliation[Also at ]{L.D. Landau Institute for Theoretical Physics,
Chernogolovka, 142432, Russia}
\affiliation{National High Magnetic Field Laboratory,
Florida State University,
1800 E. Paul Dirac Dr., Tallahassee, Florida 32310 }

\begin{abstract}
We propose that a new state with a fully gapless Fermi surface appears
in quasi-2D multiband superconductors in magnetic field applied parallel to the plane. 
It is characterized by a paramagnetic moment caused by a finite density 
of states on the open Fermi surface. We calculate thermodynamic and magnetic properties of
the gapless state for both s-wave and d-wave cases, 
and discuss the details of the 1-st order metamagnetic phase transition that accompanies 
the appearance of the new phase in s-wave superconductors. We suggest possible 
experiments to detect this state both in the s-wave (2-H NbSe$_2$) and d-wave (CeCoIn$_5$) 
superconductors.
\end{abstract}
\vspace{0.15cm}


\maketitle  

In the original BCS theory of superconductivity (SC) excited states are separated
from the ground state by an energy gap. SC does not necessarily lead to a fully 
gapped energy spectrum of quasiparticle
excitations. It is well known that for unconventional SC, where
the mechanism of SC is different from the BCS, non-trivial symmetry
of the order parameter allows the existence of point or line nodes on the Fermi
surface (FS) in momentum space\cite{VG,gorkov,Anderson}. 
For ordinary s-wave pairing gapless SC appears in the presence
of paramagnetic impurities,
when the time-reversal $t \rightarrow -t$ symmetry is broken\cite{AG}.
Neglecting the orbital effects,
for a SC placed in magnetic field one expects the realization
of inhomogeneous Larkin-Ovchinnikov-Fulde-Ferrell (LOFF) state\cite{LO,FF},
in which the superconducting gap also passes through zeros this time at the points in
real space.

A resurgence of interest to the new possibilities of gapless SC
is related to the recent proposal of ``interior gap superfluidity'' of Liu and Wilczek\cite{LW},
in which \textit{whole regions of the Fermi surface} remain ungapped. Liu and Wilczek\cite{LW}
have considered a situation often realized in atomic physics and high-energy physics\cite{gubankova}, 
and the Bose-Einstein condensation,
in which the superfluid pairing takes place for non-identical fermions condensed by an
optical trap. In this case (often called unbalanced (UB) pairing), due to both the difference in
concentrations and bare masses, the two condensing components
would have to form two FS with different Fermi momenta. For SC pairing
between the two different Fermi surfaces it is energetically favorable to 
leave a part of momentum space near the larger FS in normal state. Ground states of 
this type have been studied for ordinary SC some time ago\cite{sarma,takada}.
However, the tendency toward inhomogeneous LOFF state in ordinary s-wave SC always  
prevailed. 

In this letter we show that similar features in the energy spectrum should naturally 
appear in quasi-2D multiband SC, such as some organic SC or
``115'' heavy fermion materials, CeMIn$_5$ ($M = Co, Ir$). Most recent experimental activity 
has been devoted to finding the LOFF state\cite{KY,mizushima,kakuyanagi,mitrovic,capan} in CeCoIn$_5$. The latter
material is characterized by quasi-2D Fermi surfaces and a multiband energy spectrum.
We consider a model of quasi-2D two-band SC of both s-wave and d-wave type 
in magnetic field applied parallel to the plane. Apart from the LOFF state in higher magnetic
fields, for $\mu_B B \equiv I$ comparable with the smaller gap we observe whole
regions of open FS, similar to the situation considered in Ref.\cite{LW}. 
For an s-wave SC we investigate analytically in detail this
low-temperature and low field region of the phase diagram.

We adopt the standard multiband interaction scheme (see, e.g., Ref.\cite{ABG}).
The matrix elements  $U_{ik}(\bm{p};\bm{p'})$ for the interaction  enter the definitions of the gaps, 
$\Delta_i (\bm{p})$ for  each Fermi surface (FS) as:
\begin{equation}
\Delta_i(\bm{p}) = - T \sum_{n,k,\bm{p'}} U_{ik}(\bm{p},\bm{p'}) F_k (i \omega_n, \bm{p'}),
\label{1}
\end{equation}
where $F_k(i \omega_n, \bm{p})$ is the anomalous Gor'kov function, $k,i$ are band indices,
$U_{ik}$ is the interaction between bands $i$ and $k$. 

The type of the superconducting state below $T_c$ depends on the choice of the pairing ansatz :
\begin{equation} 
U_{ik}(\bm{p};\bm{p}') = \chi(\varphi) U_{ik} \chi(\varphi')
\label{2}
\end{equation}
where $\chi(\varphi)$ is the appropriate irreducible representation; we take $\chi(\varphi)$ 
below as a const ($1$) for the s-wave pairing,  or as $\cos{(2 \varphi)}$ for the 
d-wave pairing. Solutions of the multiband Gor'kov equations\cite{AGD,Gor:Rus} for the Green's functions in magnetic field 
$I = \mu_B B$ can be written as:
\begin{eqnarray}
\hat{F}^{\dagger}_k(\omega_n, \bm{p}) & = & 
\frac{i \hat{\sigma}^y \Delta^*_k(\bm{p})}{(i \omega_n - I\hat{\sigma}^z)^2 - \xi_k(\bm{p})^2 - |\Delta_k(\bm{p})|^2}\, \label{3} \\
\hat{G}_k(\omega_n, \bm{p}) & = & 
\frac{i \omega_n +  \xi_k(\bm{p})- I \hat{\sigma}^z }{(i \omega_n - I \hat{\sigma}^z )^2 - \xi_k(\bm{p})^2 - |\Delta_k(\bm{p})|^2}\,
\label{4} 
\end{eqnarray}
The energy spectrum of the system for excitations near each FS is given by the poles of the 
$\hat{G}_{k}(\omega_n, \bm{p})$:
\begin{equation}
 \hat{E}_k(\bm{p}) = \sqrt{\xi_k(\bm{p})^2 + |\Delta_k(\bm{p})|^2} + I \sigma^z 
\label{5}
\end{equation}
The bands with different $k$ are coupled by the gap equation, Eq.(\ref{1}). 
We consider below a model with 2 FS.

While the search for the LOFF-state commonly starts from the side of higher fields, 
we study the effects in small magnetic fields of the order of the smaller gap, $\Delta_2$. 
Our main interest lies in the field range where $\Delta_2 < I \ll \Delta_1$. 
Once the magnetic field $I=\mu_B B$ exceeds the smaller gap, 
then, according to Eq.(\ref{5}), electron- and hole- pockets will open, forming an 
ungapped area near the second Fermi surface (FS2). 
In the s-wave case this process is accompanied by a weak 1st order phase transition. 
For a d-wave SC the energy gaps $\Delta_k (\bm{p})$ have line nodes and associated gapless states from the start. 
Nevertheless, depending on the strength of interactions, an irregular behavior of the gap amplitude
as a function of magnetic field also occurs in some region of model parameters (see Fig.\ref{fig2}),
which indicates a 1-st order transition.

It is convenient\cite{ZD} to express the solution of Eq.(\ref{1}) in terms of dimensionless coupling constants, 
$\lambda_{ik} = U_{ik} \nu_k$, where $\nu_k$ is the density of states on the k-th Fermi surface (FSk). 
The linearized gap equation Eq.(\ref{1}) leads to the familiar\cite{GM} instability curve for $T_c$, which, we find,
is independent of the number of FS involved: 
\begin{equation}
\ln{\frac{T_c}{T_{c0}}\,} = \Psi\left(\frac{1}{2}\,\right) - Re\left[\Psi\left(\frac{1}{2}\, + i \frac{I}{2 \pi T_c}\, \right)\right],
\label{6}
\end{equation}
Here $T_{c0}$ is the superconducting transition temperature without the magnetic field,
\begin{eqnarray}
T_{c0} &=& \frac{2 \Lambda \gamma}{\pi}\, e^{1/g}  \ \ (g < 0) \label{7} \\
g &=& 2 \frac{\lambda_{11} \lambda_{22} - \lambda_{12} \lambda_{21}}{\lambda_{11} + 
\lambda_{22} + \sqrt{(\lambda_{11} - \lambda_{22})^2 + 4 \lambda_{12} \lambda_{21}}}\, ,
\label{8}
\end{eqnarray}
$\gamma \simeq 1.781$, and $\Lambda$  is the upper cut-off for the interactions in Eq.(\ref{2}). 
However, the total $(T,B)$ phase diagram for two bands changes significantly, especially at lower temperatures and fields. 
Some main qualitative changes in the physics of multiband SC in this area can already be seen in a simplified
model with $\Delta_1$ as the primary gap, and $\Delta_2 \ll \Delta_1$ induced by the SC order on FS1\cite{BG0}.
When $I$ is close to the primary energy gap, $\Delta_1$, an inhomogeneous LOFF state will appear\cite{vorontsov,BR}.
For $\Delta_1 \simeq \Delta_2$ there could be significant modifications for the LOFF state, and the boundaries 
of the LOFF state on the $(T,B)$ phase diagram. 
We assume that two gaps differ enough for the LOFF state not to change significantly from the single-band
model\cite{vorontsov,BR}. Below we consider in more detail the low-field region of the $(T,B)$-plane.

In the weak coupling approach, $g \ll 1$, the ratio of $\Delta_2(T,B)$, the driven gap, and $\Delta_1(T,B)$, the primary gap, which we
define as model parameter $t$, sets in at $T_c$ and is temperature- and magnetic field-independent:
\begin{equation}
\frac{\Delta_2}{\Delta_1}\, \equiv t = \frac{2 \lambda_{12}}{\lambda_{22} - \lambda_{11} + 
\sqrt{(\lambda_{11} - \lambda_{22})^2 + 4 \lambda_{12} \lambda_{21}}}\,.
\label{9}
\end{equation}

Eq.(\ref{1}) for the s-wave case can be easily solved analytically at $T=0$. 
Introducing new parameters,
\begin{equation}
\alpha = t^2 \nu_2 (\nu_1 + \nu_2 t^2)^{-1}, \ \ \Delta_0 \equiv (\pi/\gamma) T_{c0},
\label{10}
\end{equation} 
we find two different solutions for Eq.(\ref{1}) for  $I \le  I_{cr} = \Delta_0 (1 + \sqrt{1-t^2})^{-\alpha}$,
and no solutions for $I>I_{cr}$. The first solution is
\begin{eqnarray}
\Delta_1 &=& \Delta_{10} = t^{-\alpha} \Delta_0, \  I < \Delta_{20} \label{11} \\
\frac{\Delta_1}{\Delta_0}\, &=& \left(\frac{\Delta_1}{I + \sqrt{I^2 - t^2 \Delta_1^2}}\,\right)^{\alpha}, \  \Delta_{20} < I < I_{cr}. 
\label{12}
\end{eqnarray}
\begin{figure}
\includegraphics[width=3.375in]{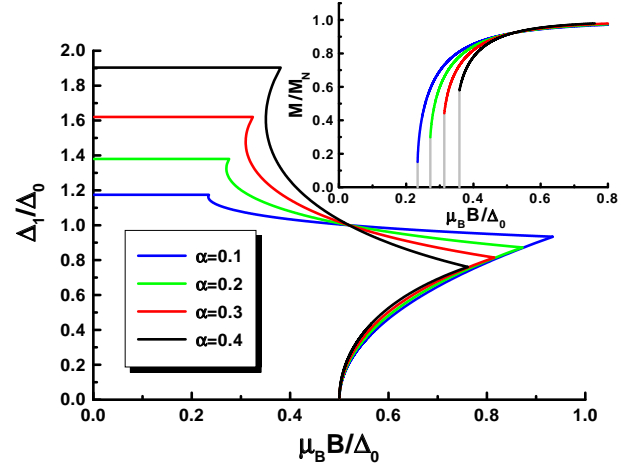}
\caption{Magnetic field dependence of the primary energy gap $\Delta_1$ for the s-wave case for $t = \Delta_2/\Delta_1 = 0.2$, 
and different values of parameter $\alpha$, $0.1$, $0.2$, $0.3$, and $0.4$. The
energy gap at $I=0$ increases with growing $\alpha$, see Eqs(\ref{9}-\ref{12}). 
The inset shows the corresponding metamagnetic transition.
The calculated paramagnetic moment is normalized to the full value of the normal moment from the second Fermi surface.
The grey lines represent the positions of the 1-st order phase transition.}
\label{fig1}
\end{figure}
Here $\Delta_{i0}=\Delta_i (T=0, B=0)$, $i=1,2$.
The second solution exists for magnetic fields $\Delta_0/2 < I < I_{cr}$,
\begin{equation}
(I + \sqrt{I^2 - \Delta_1^2})^{1- \alpha} (I + \sqrt{I^2 - t^2 \Delta_1^2})^{\alpha} = \Delta_0,
\label{13}
\end{equation}
and is the familiar\cite{Gor:Rus,LO} unstable solution for the energy gap in high magnetic fields. The two
solutions are plotted in Fig.\ref{fig1}. The re-entrant behavior in magnetic fields
$I \simeq \Delta_{20} = t \Delta_{10}$ clearly indicates the 1-st order character of transition 
into the gapless state. At this transition an open Fermi surface is formed, according to the energy spectrum
given by Eq.(\ref{5}). The position of the 1-st order phase transition is found from 
the energy at $T=0$, which for $I < \Delta_{20}$ has the usual form,
\begin{equation}
\Delta E \equiv E_S - E_{N0} = - (\nu_1 \Delta_1^2 + \nu_2 \Delta_2^2)/4. 
\label{14} 
\end{equation}
For $I > \Delta_{20}$, we also find a contribution from the normal excitations:
\begin{equation}
\Delta E = - (\nu_1 \Delta_1^2 + \nu_2 \Delta_2^2 + 2 \nu_2 I \sqrt{I^2 - \Delta_2^2})/4.
\label{15} 
\end{equation}
In Fig \ref{fig1} we also show the field dependence of the paramagnetic contribution to the total magnetization 
for the same values of the parameter $\alpha$. 
There is a characteristic metamagnetic jump in the magnetization at $I = \Delta_{20}$.
At  $I>\Delta_{20}$  the finite density of states  (DOS) arises on each new electron- and hole- FS 
that will result in the linear- in- T specific heat at low temperatures.
The 1-st order transition separating the two field regimes should be clearly seen in thermodynamic measurements.

We have obtained similar results for d-wave multiband superconductors,
such as the ``115'' materials. The general theoretical formulas involve an average over the angular
variable $\varphi$, and are more cumbersome than Eqs (\ref{11})-(\ref{15}) for the s-wave case.  
In Fig.\ref{fig2} we show, for comparison, the dependence of the amplitude of the order parameter and
the metamagnetic transition for the d-wave pairing. 
Due to the presence of line nodes in d-wave case, the energy spectrum is  gapless already at $I=0$. 
Nevertheless, similar processes take place near FS2 that could lead to a 1st order metamagnetic transition for
some values of the coupling constants. However, while for the s-wave case the 1-st order transition is always
present, for d-wave it turns into a smooth crossover for $\alpha \ll 1$.  

\begin{figure}
\includegraphics[width=3.375in]{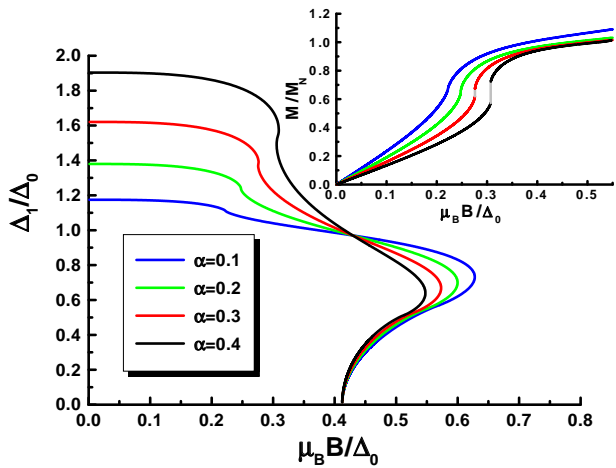}
\caption{Magnetic field dependence of the primary energy gap amplitude, $\Delta_1$, for the d-wave case for 
$t = \Delta_2/\Delta_1 = 0.2$, 
and different values of parameter $\alpha$, $0.1$, $0.2$, $0.3$, and $0.4$. The
energy gap at $I=0$ increases with growing $\alpha$.  The plot is in reduced units of $\Delta_0 = 2 \pi T_{c0}/\gamma \sqrt{e}$,
the energy gap value corresponding to $T_{c0}$ for a 1-band d-wave superconductor. 
The inset shows the corresponding metamagnetic transition,
which only occurs at $\alpha > 0.24$. The calculated paramagnetic moment is normalized to the full value of the normal moment 
from the second Fermi surface. The grey lines represent the positions of the 1-st
order phase transition.}
\label{fig2}
\end{figure}

The metamagnetic transition can be studied analytically for the s-wave case when the second gap and the
parameter $\alpha$ are small\cite{BG0}, 
$\Delta_{20} \ll \Delta_{10}$, $\alpha \ll 1$. Introducing $\tau_I \equiv (I - \Delta_{20})/\Delta_{20}$ and 
$\tau_{\Delta} \equiv (\Delta_2 - \Delta_{20})/\Delta_{20}$, we find that in the vicinity
of this transition Eq.(\ref{12}) is considerably simplified,
\begin{equation}
\tau_I = (1/2) \tau_{\Delta}^2 \alpha^{-2} + \tau_{\Delta}.
\label{16}
\end{equation}
Expanding the energy Eq.(\ref{15}) in the vicinity of this transition, we find:
\begin{equation}
E_S - E_{N0} = E_0 (1 - 2 \tau_{\Delta}^2 - (4/3) \tau_{\Delta}^3 \alpha^{-2} ),
\label{17}
\end{equation}
where $E_0$ is the energy of superconducting state at $I=0$. The cubic terms in the energy, 
and the form of Eq(\ref{16}) clearly indicate a first order transition. After a simple calculation,
we find that the 1-st order transition occurs at $\tau_{Icr} = - 3 \alpha^2/8$. The
energy gap $\tau_{\Delta}$ changes abruptly from $\tau_{\Delta} = 0$ to $\tau_{\Delta} = - 3 \alpha^2/2$,
which corresponds to a metamagnetic transition, with a paramagnetic jump in the magnetic moment 
$M = (3 \alpha/2) \mu_B \nu_2 \Delta_{20}$ (or to a sudden appearance of the finite DOS). We find
a simple expression for the magnetic moment for $I > I_{cr}$ in the vicinity of this transition:
\begin{equation} 
M = \mu_B \nu_2 \Delta_{20} (\alpha + \sqrt{\alpha^2 + 2 \tau_I}).
\label{18}
\end{equation}

The paramagnetic moment in the s-wave case always appears at $I \gtrsim \Delta_{20}$ in a 1-st order phase transition.
Nevertheless,
in case of the driven second gap, $\Delta_{20} \ll \Delta_{10}$, the 1-st order transition is weak, and so
is the corresponding change in the SC order parameter, $\Delta_2 = t \Delta_1$. In the first 
approximation this change can be neglected\cite{BG0}. Then the temperature- and field- dependence
of the magnetic moment is completely described by the standard\cite{mineev} formulas that follow from
the energy spectrum Eq.(\ref{5}), where $\Delta_2 = \Delta_{20}$ is regarded as a constant. For example, a simple analytic
expression for the magnetic moment in the s-wave case at $T=0$ is:
\begin{equation}
M = \mu_B \nu_2 \sqrt{I^2 - \Delta_2^2}, \ \ I > \Delta_2. 
\label{19}
\end{equation}
Note that $M=0$ for $I < \Delta_2$. For a d-wave case, FS2 gives a contribution
for $I$ above and below $\Delta_2$:
\begin{equation}
M_2 = \frac{2}{\pi}\, \mu_B \nu_2 \int_0^A d \varphi \sqrt{I^2 - \Delta_2^2 \sin^2 \varphi},
\label{20}
\end{equation}
where the upper limit is $A=\pi/2$ for $I > \Delta_2$, or $A = arcsin(I/\Delta_2)$ for $I<\Delta_2$.
This is an elliptic integral of second kind. Note that in the d-wave case FS1 gives the usual nodal
contribution, $M_1 = 0.5 \mu_B \nu_1 I^2/\Delta_1$. The density of states for FS2 in the s-wave case
is also given by a simple formula:
\begin{equation}
\nu_2(I) = \nu_2 I/\sqrt{I^2 - \Delta_2^2}.
\end{equation}
In the d-wave case one has to introduce the appropriate angular averages of this result, and add
the familiar nodal contribution from the first Fermi surface\cite{mineev}.

In summary, we have shown that a gapless Fermi spectrum characterized by open Fermi surfaces is an inevitable feature 
for a quasi-2D multiband superconductor placed into a large enough field  parallel to the plane. 
The new state is fully analogous to the one studied in Ref.\cite{LW} for the 
unbalanced pairing problem. Unlike Ref.\cite{LW}, however,  such a  gapless state sets in 
as the 1-st order transition in increased magnetic field. 
Measurements of the specific heat in applied field are the most direct way to observe the effect 
in s-wave superconductors, such as 2H-NbSe$_2$\cite{yokoya,sologubenko}.
The transition also leads to a metamagnetic jump in 
the magnetization. For a d- wave pairing, because of the nodes, gapless excitations are present even 
without external field. As the field is increased, the open Fermi surfaces develop gradually, 
although character of the process may depend on the interaction parameters. Applied fields should be 
low enough for these phenomena not to interact with the LOFF state.
It is broadly believed\cite{KY,mizushima,capan} that the properties of CeCoIn$_5$  may be close enough to a two-dimensional 
model to display the inhomogeneous LOFF state\cite{comment}. If CeCoIn$_5$,indeed, belongs to the strongly quasi-2D class, 
the low field properties studied above should 
manifest themselves as well. 
Then, if interactions in CeCoIn$_5$ were strong enough to result in a 1-st order transition of Fig.\ref{fig2}, the latter could 
be observed best by calorimetric measurements, as for the s-wave pairing. If not, then one may rely on the NMR methods 
for the observation of a rather non-monotonic field behavior for non-linear susceptibility shown in the insert of Fig\ref{fig2}
(we have not considered possible implications of the effect for thermal conductivity in the presence of a magnetic field).  
The above effects should be expected for other 2D organic compounds. An ideal realization of the scheme would be 
superconductivity localized at the surface\cite{BG}.

The authors are thankful to L. Balicas and Z. Fisk for helpful discussions.
This work was supported (VB) by TAML at the University of Tennessee and (LPG) by NHFML through
the NSF Cooperative agreement No. DMR-008473 and the State of Florida.


\begin{thebibliography}{}
\bibitem{comment} In the real situation of a Q2D material the carriers "inside the plane" , in addition to 
the external field, would, of course, experience a contribution from the inhomogeneous  
field of penetrating vortices.
\end{thebibliography}
\end{document}